# Reconstruction of three-dimensional strain field in an asymmetrical curved core-shell hetero-nanowire

Serhii Kryvyi[1,*], Slawomir Kret[1] and Piotr Wojnar[1]

[1]Institute of Physics Polish Academy of Sciences, al. Lotnikow 32/46, 02-668 Warsaw, Poland

E-mail: kryvyi@ifpan.edu.pl

**Abstract**

Crystal orientation and strain mapping of an individual curved and asymmetrical core-shell hetero-nanowire is performed based on transmission electron microscopy. It relies on a comprehensive analysis of scanning nanobeam electron diffraction data obtained for 1.3 nm electron probe size. The proposed approach handles also the problem of appearing twinning defects on diffraction patterns and allows for investigation of materials with high defect densities. On the basis of the experimental maps and their comparison to finite element simulations, a hidden core-shell geometry and full three-dimensional strain distribution within the curved core-shell nanowire are obtained. As effect, a low-dose quasi-tomography data using only single zone axis diffraction experiment is received. Our approach is applicable also for electron beam sensitive materials for which performing conventional tomography is a difficult task.

**Headlines:**

*Unique approach for strain and orientation mapping within a single nano-object with defects*

*Strain and orientation mapping of single curved highly-asymmetric core-shell nanowire*

*A detailed description of the method for 3D strain reconstruction in core-shell nanowire based on single zone axis diffraction pattern and its comparison to finite element method simulation*

# 1. Introduction

Epitaxial hetero-nanowires (NWs) represent a novel technological platform with a wide range of applications in nanotechnology, covering a number of electronic and photonic devices, as well as sensors [1]. Among others, the core-shell NWs have attracted a great deal of attention for their applications in future nanodevices. Implementation of core-shell geometry allows for the enhancement of the light extraction efficiency from light-emitting diodes (LEDs), a significant increase of light absorption in solar cells, and an effective electric field control of electric current in transistors resulting in a considerable improvement of the device performance [2]–[4]. The presence of the shell may additionally result in the enhancement of optical emission intensity from the NW-cores, as demonstrated for GaAs/(Ga,In)P [5] and (Zn,Mn)Te/(Zn,Mg)Te core/shell nanowires [6].

Moreover, the core-shell NWs are promising building block for nanoelectromechanical systems (NEMS) [7],[8]. However, due to the lattice mismatch between the core and the shell semiconductor, the core-shell configuration introduces a relatively complex anisotropic three-dimensional (3D) lattice distortion in contrast to two-dimensional (2D) tetragonal distortion typical for thin epilayers deposited on a thick lattice-mismatched substrate. This principally different behavior originates from the relatively small lateral dimensions of the NWs and the large surface-to-volume ratio typical for these structures.

For axial NW heterostructures, it is well known that the critical thickness increases with decreasing NW's radius [9]. In the case of radial NW heterostructures the situation is more complex. In particular, the contribution of surface and interface energy significantly affects the strain relaxation mechanism in core-shell NW heterostructures. Therefore, it is more appropriate to use the concept of critical dimensions instead of only one critical thickness. The NW critical dimensions: NW length, core radius, and shell thickness, depend on each other [10]. Moreover, they are also affected by NW crystallographic orientation and the presence of facets.

Significant strain fields that occur in lattice mismatched core-shell NWs can cause the formation of lattice defects, misfit dislocation in most cases, and surface morphological transformations, such as stress-driven surface roughening [11]–[13]. In addition, the influence of the core cross-section shape on the strain field in core-shell NWs has been reported [14]. Importantly, core-shell NWs are very sensitive to radial inhomogeneity. Even small asymmetry of the shell composition or NW-core decentering leads to the occurrence of noticeable asymmetric strain fields and, as result, bending of the NW. This bending reduces stresses in the NW heterostructure and, thus, leads to the avoidance of defect formation.

Strain fields that inevitably occur in lattice mismatched core-shell NWs have a significant impact on their optoelectronic properties and, in spite of the potential formation of defects, can be used to fine-tune effectively the band gap energy [15]–[18]. Moreover, the radial core-shell geometry of the NW heterostructures gives rise to the realization of a number of crystalline coherent structures unobtainable in classic planar configuration [19],[20]. These findings motivate the investigation of strain fields in core-shell hetero-NWs. Understanding and prediction of the strain relaxation mechanism is crucial for the development of novel and the improvement of the quality of existing hetero-NW based devices. Therefore, a significant number of papers is dedicated to the study of strain distribution within core-shell NWs has appeared [13],[21]–[36].

Among different experimental techniques, only X-ray synchrotron and transmission electron microscopy (TEM) based methods provide an opportunity for strain mapping with a sufficiently high spatial resolution. Our previous report has shown that using scanning nanobeam electron diffraction (NBED) in TEM allows us to overcome the difficulties related to NW thickness variations and twinning of a crystal structure [37] and leads to the realization of precise 2D strain maps of axial NW heterostructures. Our approach does not employ the precession of electron beam, but it makes possible the conservation of dynamical effects on diffraction pattern (DP) during acquisition of experimental data. The dynamical features in diffraction disks, for instance, can be effectively used for specimen thickness determination with nanometer precision [38],[39].

The purpose of this study is to use the intensity distribution of reflections together with their position for the realization of crystal orientation maps in addition to strain maps. To achieve this goal, the so-called template matching technique is applied. It relies on the comparison of each experimental diffraction pattern to the database of pre-simulated dynamical DPs for different misorientation angles. As an effect, the crystalline orientation mapping is performed with unusually high resolution. Those results allow us to reveal quasi-tomography data of a core-shell NW and to conduct a full reconstruction of the strain field within an individual complex nano-object.

A similar approach described previously used the commercial ASTAR software (by Nanomegas) to recognize the phase and crystal orientation in polycrystalline materials [40]. In that case the kinematical electron diffraction approximation has been employed to generate the template database. A recent paper reported using a dynamic approach for orientation mapping [41]. However, those approaches rely on the precession of electron beam for collecting of experimental data. Those limitations impair the angular resolution for the crystal orientation mapping. While a typical angular resolution of 1 degree is well enough for a polycrystalline sample it is not sufficient for a monocrystalline object which is subject of our present study.

For the investigation of a single monocrystalline NW the use of X-ray micro Laue diffraction [42] and selected area electron diffraction (SAED) [43] for local lattice orientation and strain profiling along the NW axis have been reported. Both approaches are characterized by an order of magnitude better angular resolution for local lattice orientation as compared to the previously mentioned kinematical approach [40], but the spatial resolution remains poor. Therefore, a new approach is required to map small angles of misorientation for single-crystal objects.

In this work, we study the strain distribution in radial (core-shell) NW heterostructures by a conventional TEM technique. NBED is used for the relative lattice mapping with picometer precision and a few nanometers spatial resolution. In addition to the lattice mapping, the pattern matching technique provides additional information about the sample such as its thickness and local crystallographic orientation. The proposed approach allows us, therefore, to obtain local crystallographic orientation mapping with ultimate precision up to 0.1 degree. The obtained data are used to create realistic finite element method (FEM) models to perform full three dimensional tomography reconstruction of strain field in radial NW heterostructure.

## 2. Methods

The ZnTe/(Cd,Zn)Te core-shell NWs were grown on (111)-oriented Si substrate by molecular beam epitaxy (EPI 620 system) with the use of Au nano-catalysts employing the vapor–liquid–solid (VLS) growth mechanism. 1-nm-thick gold layer was pre-deposited on the Si substrate and heated up to 450°C to induce the formation of Au-Si liquid nano-droplets which serve as catalysts for the further ZnTe NWs growth. (Cd,Zn)Te shells are deposited after the reduction of the substrate temperature to 350°C, at which the catalyst droplets are frozen leading to the epitaxial overgrowth of ZnTe NWs. The detailed growth procedure is almost the same as described in Ref. [44].

Structural investigations and elemental composition characterization were performed using an objective lens corrected FEI Titan Cubed 80-300 microscope operating at 300 kV. For the TEM examination, the NWs were transferred mechanically by sliding a copper grid covered with holey carbon film across the sample surface with NWs.

High-resolution transmission microscopy (HR-TEM) images were acquired by Gatan Ultrascan 1000 charge coupled device (CCD) camera. For high-resolution scanning transmission electron microscopy (HR-STEM) in annular dark field mode, the high-angle annular dark-field (HAADF) Fischione 3000 detector was used. Nanobeam electron diffraction (NBED) mapping of (Cd,Zn)Te/ZnTe hetero-NW was performed in STEM mode using the Gatan Ultrascan 1000 camera with binning 4 to increase the dynamic range and to reduce a blooming effect. Diffraction patterns were acquired with a convergence semi-angle of 1 mrad and a probe size of 1.3 nm [37]. The integration time for NBED acquisition was chosen to be 100 msec per frame. Energy dispersive X-ray spectroscopy (EDX) data were collected in TEM by EDAX 30 mm$^2$ Si(Li) detector with a collection angle of 0.13 srad.

The analysis of NBED data was performed by a scripting procedure using DigitalMicrograph software. To obtain precise strain maps with subpixel accuracy the circular Hough transform was implemented in the script [37],[45]. The pattern matching technique to correlate experimental diffraction patterns and simulated data was used to obtain local orientation maps of bent NW. Since NBED data was collected for [110] zone axis, the orientation mapping in our case is based on the determination of crystallographic misorientation of the NW with respect to [110] zone axis. The procedure is based on a cross-correlation algorithm and is also implemented in the DigitalMicrograph script. The simulation of tilt series was performed using QSTEM software which is based on a dynamical approach and a multislice algorithm [46]. A set of misoriented DPs was simulated for [110] ZnTe crystal misoriented in in the range from -4 to 4 degrees towards [001] and [1$\bar{1}$0] crystallographic directions with step of 0.02 degrees. These tilt series were simulated for crystal thicknesses up to 80 nm with the step of 1 nm.

The finite element method (FEM) modeling of (Cd,Zn)Te/ZnTe core-shell NW was performed by COMSOL Multiphysics software. The 3D model was developed based on the measured geometric characteristics of NW. The model consists of ZnTe NW-core surrounded by an asymmetric (Cd,Zn)Te shell. Crystallographic orientation and anisotropy of Young modulus [37] were also taken into account in the model. The simulated NW has the shape of a hexagonal frustum with the height of 235 nm and the length of the bottom face edge of 30 nm and that of the top face of 25 nm. The last two numbers can be roughly considered as radii of the NW on the bottom and top bases, respectively. To eliminate surface relaxation effects on the bottom and top NW bases additional segments with 50 nm height were added to these bases. These additional parts were not considered in the analysis of the simulated data.

## 3. Structural studies

EDX elemental mapping of a typical ZnTe/(Cd,Zn)Te NW segment confirms the radial character of the NW heterostructure. However, an asymmetry in the shell thickness occurs, which is confirmed by the irregular radial distribution of cadmium across the measured NW heterostructure (Fig. 1(b)). The average Cd content, x, in $Cd_xZn_{1-x}Te$ shell determined by EDX is about 0.6±0.05. In the core-shell configuration the asymmetrical shell and the relatively large lattice misfit between ZnTe core and (Cd,Zn)Te shell leads to a significant bending of the NW (Fig. 1(a)). As a result, atomic columns within this NW heterostructure are also bent having as a consequence that the standard geometrical phase analysis (GPA) method is not applicable for monitoring lattice changes on a large fragment of this sample. Nevertheless, the HR-TEM/STEM investigations evidenced an epitaxial growth of the (Cd,Zn)Te shell on the ZnTe core. No misfit dislocations between the core and the shell were observed, whereas the crystalline structure of the shell is exactly the same as the crystalline structure of the NW core. Twin boundaries propagating from the core to the shell without any discontinuity were observed (Fig. 1(b)). In addition, a thin 4-5 nm amorphous layer on the NW surface is present. We associate it to the oxidation of NWs in the air, which is confirmed by the increased oxygen content on the surface found in EDX data.

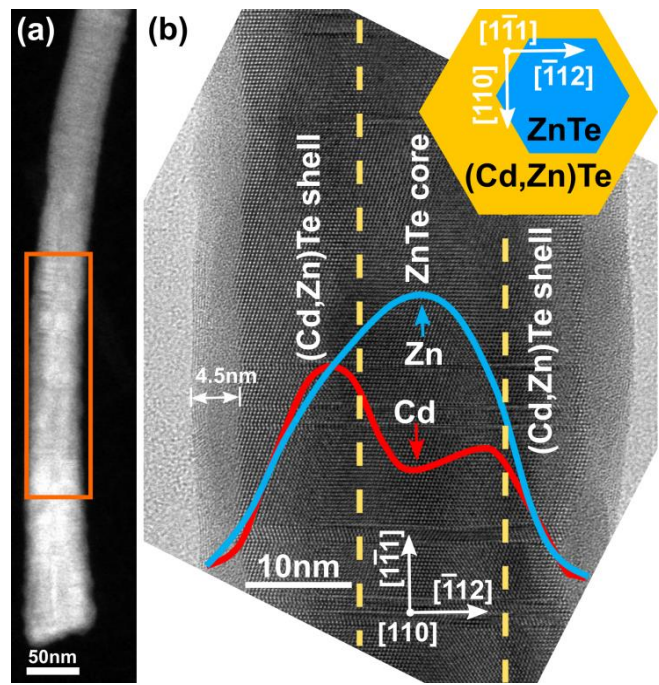

**Figure 1.** (a) The High Angle Annular Dark Field image of ZnTe/(Cd,Zn)Te core-shell NW measured in [110] zone axis; the rectangle represents an area investigated by nanobeam electron diffraction (NBED). (b) HR-TEM image of a typical NW segment; dashed lines indicate the hypothetical position of ZnTe core. The red and blue line profiles represent the intensities of $Cd_L$ and $Zn_K$ lines from EDX spectra averaged across the NW, respectively. The inset represents the schematic image of an asymmetric core-shell NW cross-section.

## 4. Strain mapping

Diffraction patterns for (Cd,Zn)Te/ZnTe radial NW heterostructure were acquired as a grid of scans recorded point by point, line by line, and finally as a 3D stack. The acquisition area with a sampling of 24 px in the radial and 60 px in the axial direction of the NW was used. The pixel size was 2.82 nm and 3.91 nm for x and y directions, respectively, leading to the total area with dimensions 67.6×235 nm. Diffraction patterns were recorded for each electron beam position, thus, the final experimental stack consists of 1440 diffraction patterns.

NBED was used to analyze the local crystal lattice deformation, i.e., to perform the strain mapping of the core-shell NW heterostructure. For this reason, the scripting procedure written in DigitalMicrograph software was used. This procedure consists of finding the distance $\boldsymbol{g}_{hkl}$ between (000) disk and a diffraction disk with (*hkl*) indexes. The choice of disk that corresponds to diffraction from a certain crystal plane allows for mapping of the interplanar spacing for specific crystallographic directions. Typically, the relative strain is introduced into this procedure:

$$\varepsilon_{hkl}^{rel} = \frac{d_{hkl}^{exp} - d_{hkl}^{ref}}{d_{hkl}^{ref}} \qquad (1)$$

where $d_{hkl}^{exp}$ and $d_{hkl}^{ref}$ are an experimental and reference interplanar spacing value for (*hkl*) lattice planes respectively. Or in terms of reciprocal space, using $|\boldsymbol{g}_{hkl}| \sim \frac{1}{d_{hkl}}$, Eq. (1) can be written as:

$$\varepsilon_{hkl}^{rel} = \frac{|\boldsymbol{g}_{hkl}^{ref}| - |\boldsymbol{g}_{hkl}^{exp}|}{|\boldsymbol{g}_{hkl}^{exp}|} \qquad (2)$$

where $\boldsymbol{g}_{hkl}^{exp}$ and $\boldsymbol{g}_{hkl}^{ref}$ are an experimental and reference distance between (000) and (*hkl*) disk, respectively. As the reference, an experimentally obtained value is usually used. As a result, a strain map represents lattice change with respect to a reference area on the map. In our considerations, we have chosen an area with maximal $\boldsymbol{g}$ value (i.e. minimal value of interplanar spacing) as the reference. However, in the case of core-shell NW, it also may be useful to use other reference values. For example, the absolute strain $\varepsilon^{abs}$ can be calculated using the relaxed value of core and shell lattice as a reference for the core and shell region, respectively.

It should be noted, that information obtained by NBED is averaged along the electron nanobeam path. That is why the strain maps represent projections of lattice change on the image plane. Nevertheless, maps can also be a function of electron probe defocus. In our case however, the latter effect can be neglected due to the large depth of field typical 1 mrad convergent probe. This significantly simplifies strain appearance represented on maps compared to complicated real 3D behaviors of the lattice plane in asymmetric core-shell NW.

Prior to performing the strain mapping, one has to link the crystallographic directions and the geometry of the NW (Fig. 2(a)). As can be seen in Fig. 2(a), the identification of $(2\bar{2}\bar{4})$ and $(1\bar{1}1)$ reflections lead us to determine directly radial $\varepsilon_r^{rel}$ and axial $\varepsilon_a^{rel}$ strain in NW, respectively. Please note, that in the case of our measurements we have the insight into the strain component perpendicular to the electron beam only. That is why the radial strain $\varepsilon_r^{rel}$ in our considerations is restricted to this component only.

Importantly, a large number of crystal twins typical for (Cd,Zn)Te structure makes it difficult to analyze NBED data. According to analyzed NBED data, several types of DP can be found on the experimental stack. The first one is a regular pattern, as shown in Fig. 2(a), there are no difficulties to analyze such DP. The second type represents a pattern in which the diffraction from both twins appears simultaneously (Fig. 2(b)). In this case the electron probe reaches a region that is close to the twin boundary.

The situation is more complex in the case of the other three types. Additional disks that cannot be associated with any of the twin crystal lattice appear on DPs (Fig. 2(c-e)). For instance, in the third type of DP only reflections on the line in the growth direction and on the same line as the $(2\bar{2}\bar{4})$ and $(1\bar{1}1)$ reflections appear and are well separated from each other (Fig. 2(c)).

In a fourth type of DP, an additional disks generation arises for each reflection on the line of reflections corresponding to the growth direction (Fig. 2(d)). Nevertheless, the recognition of individual reflections is still possible for this type of DP. The last type of DPs, presented in Fig. 2(e), is the most difficult for the analysis. This is due to the overlapping and smearing of the major part of the disks. This applies especially to a series of (111) reflections. This effect significantly limits the number of reflections that can be used for strain mapping.

Based on the appearance of different types of DPs and on the positions of the particular reflections on the DPs a map of twins distribution of the investigated NW segment is obtained (Fig. 2(k)). The blue and brown regions on the map correspond to the twin A and B, respectively (Fig. 2(i,j)). The orange color on this image refers to the area that cannot be described to any of the twins and belongs to one of the types of DPs presented on the Fig. 2(b-e). This map presented in Fig. 2(k) is used in our further considerations, in the algorithm for the determination of the crystal orientation.

In spite of the complexity of the data, it is possible to determine the positions of diffraction disks and, as result, to perform the strain mapping. For this purpose we use $(2\bar{2}\bar{4})$ and $(1\bar{1}1)$ reflections. In addition to the fact, that those reflections allow the direct insight into the strain in the radial and axial direction, they are common for both twins types and can be well resolved even for the "intermixed" DPs.

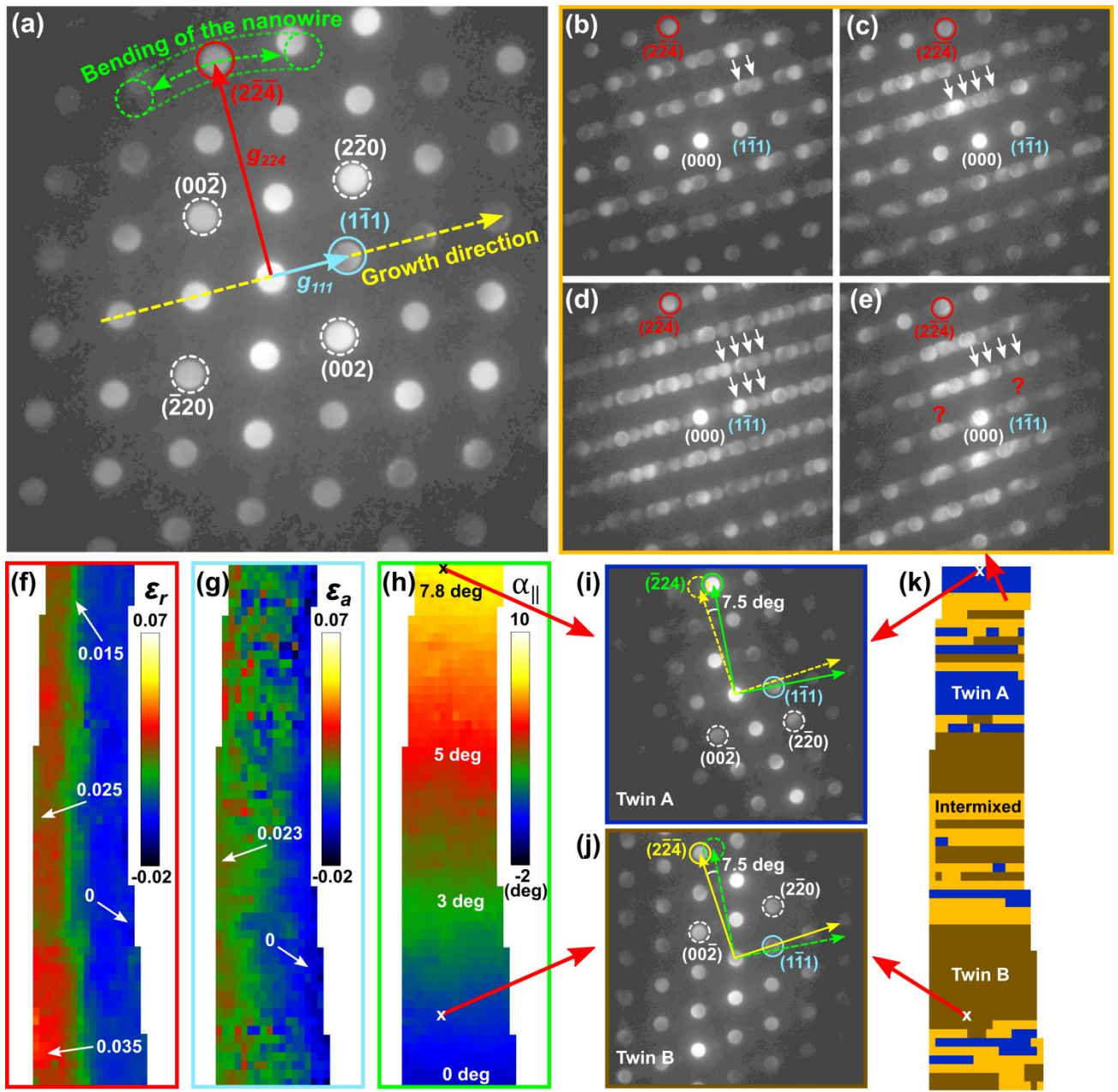

**Figure 2** Results of NBED analysis of core-shell ZnTe/(Cd,Zn)Te NW. (a) Typical nanobeam electron diffraction pattern with indicated growth direction and some of the reflections. (b-d) Four types difficult to analyze diffraction patterns were found in the experimental data. (f, g) Maps of relative lattice change are calculated by using $(2\bar{2}\bar{4})$ reflection for radial and $(1\bar{1}1)$ reflection for axial direction respectively. (h) The map of in-plane component of NW's bending received using $(2\bar{2}\bar{4})$ reflection. (i, j) NBED patterns correspond to twin A and twin B respectively. (k) Map of the twins distribution in the NW. For intermixed regions features corresponding to both twins A and B appear simultaneously.

The results of strain mapping reveal a strong radial asymmetry of the lattice parameters (Fig. 2(f) and 2(g)). This observation is consistent with EDX data demonstrating an inhomogeneous (Cd,Zn)Te shell thickness. There are no significant discontinuities or strain jumps in the NW axial direction visible on the maps associated to the presence of twins. On the other hand, the radial strain map $\varepsilon_r^{rel}$ representing lattice change in the direction perpendicular to the NW axis, shown in Fig. 2(f), reveals a region of a distinct increase of the strain up to 3-3.5%. This strain step occurs in the radial direction of the NW and is present along the NW growth axis within the entire investigated NW segment. The region characterized by strain jump corresponds well to the boundary between the core and shell. Therefore, the above described $\varepsilon_r^{rel}$ - map can be used to estimate the diameter and position of NW-core with respect to the NW center. The presence of only one strain jump on the $\varepsilon_r^{rel}$ map reflects the highly asymmetric nature of the shell thickness. It can be concluded that the almost entire shell is present on the left side of the NW while on the right side either bare ZnTe or a significantly thinner (Cd,Zn)Te shell is present.

The axial $\varepsilon_a^{rel}$ strain map shown in Fig. 2(g) is characterized by a certain noise. There are two reasons for this: the first one is the smaller value of **g** vector for $(1\bar{1}1)$ reflection compared to $(2\bar{2}\bar{4})$ and, as result, the larger impact of uncertainties related to the determination of its position on the DP. The second reason is related to the difficulties of the determination the $(1\bar{1}1)$ disk position on some DPs, e.g., on DPs shown in the Fig. 2(e). Despite of the presence of some noise, a relatively smooth lattice change is clearly visible on the map shown in Fig. 2(g). We attribute this lattice distortion to the bending of the NW since the core and shell crystal lattices are coherent in the axial direction and no strain in the axial direction is expected.

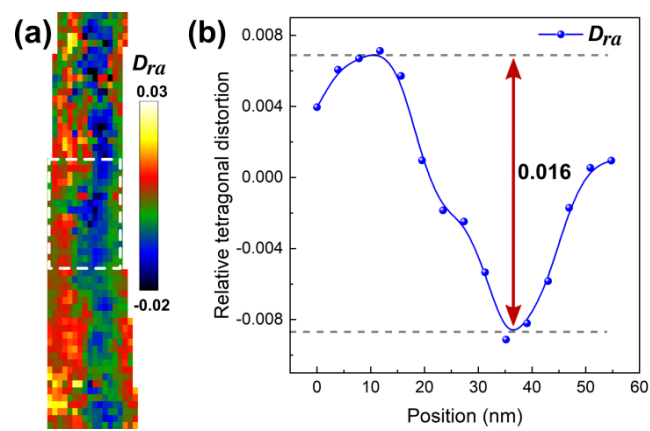

**Figure 3** (a) Lattice distortion map calculated according to Eq. (3). (b) The average profile calculated for the dashed region on (a)

According to the data presented in Fig. 2, slightly larger strain values are observed for the radial than for the axial strain. This means that a variation of lattice parameters in the NW is larger for radial direction as compared to axial direction. This result can be understood in terms of a quasi 2D misfit strain acting on the shell in the core-shell geometry [32]. In order to characterize the value of the lattice distortion, the following simple equation describing the tetragonal deformation is used:

$$D_{ra} = \varepsilon_r - \varepsilon_a \qquad (3)$$

where $D_{ra}$ is the lattice distortion, $\varepsilon_r$ and $\varepsilon_a$ are radial and axial strain respectively. The lattice distortion map calculated by Eq. (3) is presented in Fig. 3(a), where the core-shell geometry of the NW is clearly reproduced. It should be noted that the values in Fig. 3 represent relative lattice distortion with respect to a reference region, and $D_{ra} = 0$ doesn't necessarily correspond to the lattice without tetragonal deformation. Nevertheless, the results presented in Fig. 3 provide information about the character and range of the lattice distortion. According to the average profile presented in Fig. 3(b) determined for the dashed region of the distortion map (Fig. 3(a)), the relative lattice distortion reaches values up to 1.6%. The dependence of relative tetragonal distortion on the radial position of the nanowire can be well reproduced by the calculations in terms of the elastic theory (see the Electronic Supplementary Material (ESM) Fig. S1).

In addition to the strain maps, also the map of the in-plane component (in-plane of view) for lattice bending $\alpha_\parallel$ is obtained (Fig. 2(h)) based on diffraction patterns. To determine this parameter the change in azimuthal angular position of $(2\bar{2}\bar{4})$ disk with respect to $(000)$ beam is measured (Fig. 2(a,i,j)). Although $\alpha_\parallel$ value reflects the bending of $(2\bar{2}\bar{4})$ crystal plane it can also be applied for the description of the bending of the $[1\bar{1}1]$ NW growth axis. The choice of the $(2\bar{2}\bar{4})$ reflection for this purpose is explained by the error minimization. For the investigated 235 nm long NW segment the in-plane NW bending of about 8 degrees is observed (Fig. 2(h)). The result agrees well with STEM data, where NW's bending is also clearly visible.

To summarize, based only on the diffraction disks positions on NBED pattern, it is possible to calculate relative axial and radial strain components, quasi tetragonal lattice distortion, and the in-plane components of the lattice bending. However, additional information can be obtained from the analysis of intensity redistribution within the DP and within individual diffraction disks.

## 5. Orientation mapping

This section describes a procedure for mapping of the local crystal orientation in three dimensions (3D). For convenience, according to the Cartesian coordinate system, three bend angles $\alpha_\parallel$, $\alpha_\perp$, $\alpha_t$ are chosen to characterize the local crystallographic orientation of the NW. The in-plane component of lattice bending $\alpha_\parallel$ was determined in the previous section based on the in-plane rotation of DPs by following the position of the $(2\bar{2}\bar{4})$ reflection. This component represents an angle of unit cell rotation in the plane perpendicular to the electron probe. The other two angles $\alpha_\perp$, $\alpha_t$ correspond to the bending components toward [1$\bar{1}$1] growth axis and angle of twist of the crystal lattice around [1$\bar{1}$1] axis (Fig. 4), respectively. The last two components are similar to $\alpha$ and $\beta$ tilt angles that are used in TEM for specimen orientation and can be also named *off-axis angles*.

All bend angles are determined relative to some reference values. As the reference for $\alpha_\parallel$ the crystal orientation of most bottom part of the NW is chosen. In the case of $\alpha_\perp$ and $\alpha_t$ the perfect [110] zone axis orientation is chosen as the reference.

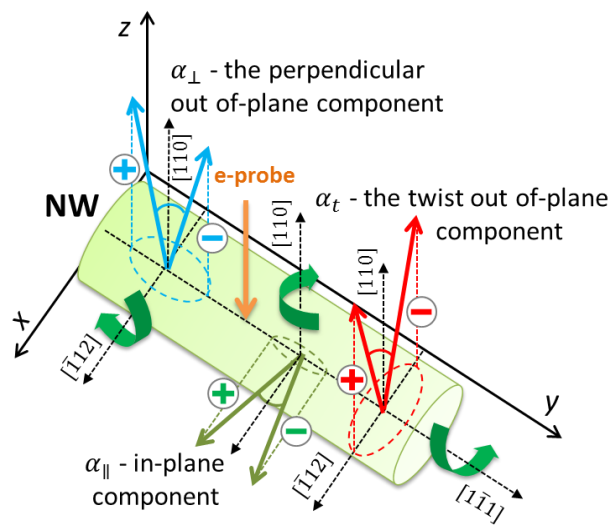

**Figure 4.** The scheme represents the definition of the misorientation angles. $\alpha_\parallel$ – angle denotes the in-plane component of lattice bending and $\alpha_\perp$, $\alpha_t$ – out-of-plane components. The sign "+" and "-" represents the bend direction.

While the in-plane component of lattice bending ($\alpha_\parallel$) causes the rotation of the diffraction patterns, the out-of-plane component leads to the redistribution the reflections intensity on diffraction patterns (Fig. 5(a-d)). However, the effect of the Ewald sphere intersects the reciprocal lattice, which is sensitive to both, sample tilt and thickness. This fact is used for a detailed study of the experimental diffraction data and to obtain local crystal orientation maps of the investigated object. The idea of the determination of the crystal orientation at a given position of the electron beam relies on the comparison of the experimental data to a set of simulated patterns for various tilt angles.

For a given material there is only one crystal orientation near the zone axis that corresponds to a specific diffraction pattern. That is why a full pattern matching technique assisted with a cross-correlation procedure can be used for the determination of the off-axis angles $\alpha_\perp$ and $\alpha_t$ for each experimental point based on the cross-correlation maps, as exemplified in Fig. 5(e). However, some geometric transformations are required prior to the application of pattern matching technique. They require the previously calculated strain, in-plane orientation and twinning maps. Transformation of experimental patterns (instead a simulated one) was used for decreasing computation time. In the case the twin type cannot be clearly defined (intermixing region on the twinning map) the pattern matching procedure is performed for both twins. Then, angles $\alpha_\perp$ and $\alpha_t$ for the twin with a higher correlation index are further considered.

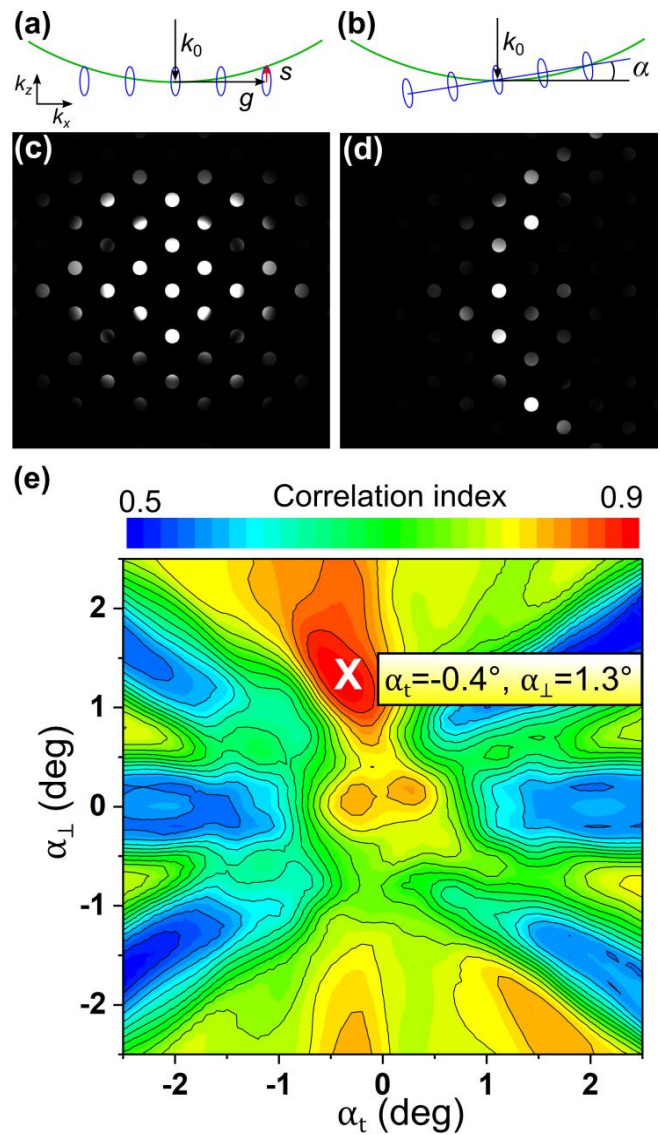

**Figure 5.** (a), (b) – Regions of reciprocal space close to (000) for on-axis and off-axis conditions, respectively. The diffraction pattern is formed due to the intersection of the relrods by the Ewald sphere (c), (d). (a) – zone axis condition provides a quite uniform intensity distribution between diffraction disks as shown on (c). (b) – off-axis condition causes the vanishing of some diffraction disks as shown on (d). (e) – exemplary correlation map for one of the experimental diffraction patterns.

As mentioned above, the intensity distribution on DP for a given crystal depends not only on crystal orientation but also on the crystal thickness. Therefore, for the precise crystal orientation mapping the pattern matching procedure is performed for various simulated crystal thicknesses. As a result, 3D stacks of correlation maps are obtained. Each (*x,y,z*) position on the stack corresponds to off-axis angles $\alpha_\perp$, $\alpha_t$ and crystal thickness, respectively. Thus, along with mapping $\alpha_\perp$, $\alpha_t$ it is also possible to estimate the thickness (t) of the NW at each experimental point. To perform a more precise crystal thickness mapping a relatively large electron probe converge angle would be required. This would, however, significantly complicate the diffraction disks recognition and, as result, worsen the strain mapping.

The results of crystal orientation maps and the map of the estimated NW thickness are presented in Fig. 6. Despite some noise in the data, a clear tendency is found on each map. According to the definition of the off-axis angles, the region of NW for which $\alpha_\perp$ and $\alpha_t$ are equal 0 corresponds to [110] zone axis orientation.

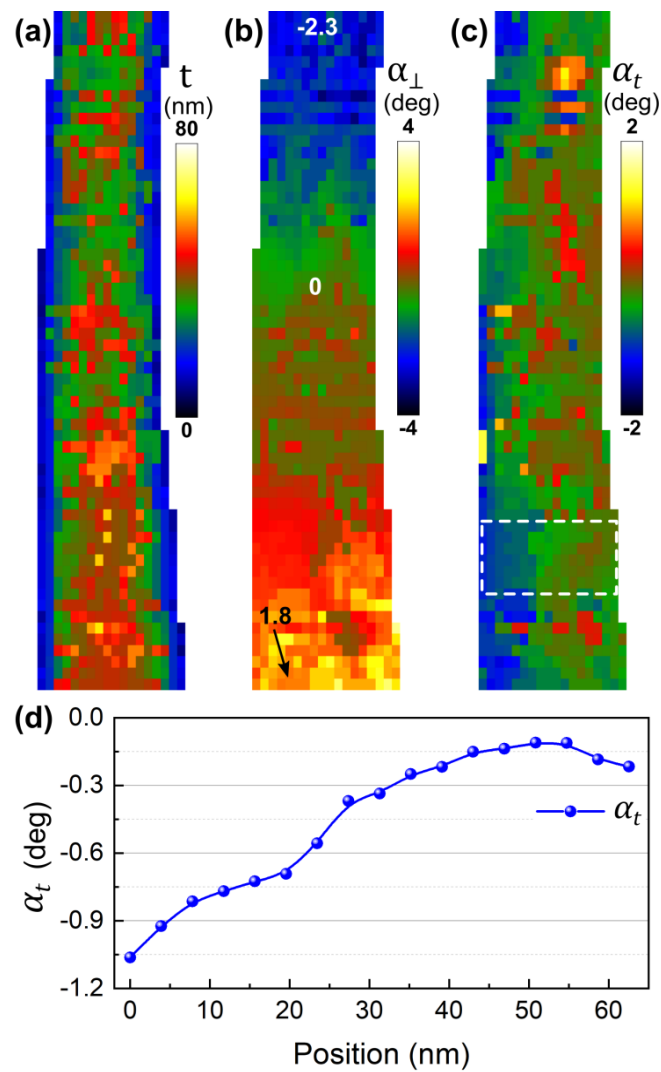

**Figure 6** The results of the pattern matching procedure: (a) - thickness map. (b), (c) - maps of out-of-plane components of bending $\alpha_\perp$ and twisting components $\alpha_t$ respectively. (d) The average profile of $\alpha_t$ calculated for the dashed region on (c)

It is worth reminding that all maps represent a projection on the plane of view which is important for their interpretation. That is why we may conclude that according to results presented in Fig. 6 the top part of investigated NW's fragment is bent with respect to bottom part of about 4.1 degrees in the direction perpendicular to the plane of view. The "minus" sign on the $\alpha_\perp$ map defines the direction of the detected bending according to Fig. 4 and points that the NW is bent in the direction of the electron beam propagation. Special attention should be paid to the out-of-plane twisting component map $\alpha_t$, Fig. 6(c). This map represents the average value of (110) planes tilt in the radial direction of the NW (or the twist [1$\bar{1}$1] growth direction). A negative value is interpreted as tilting of (110) planes to the left, with respect to the position of the NW shown in Fig. 6. The averaged profile of $\alpha_t$ for the least noisy area on the map is presented in Fig. 6(d). Interestingly, this value is not constant and has the tendency to change in the radial direction. This observation means that a complex lattice distortion is present in the asymmetric core-shell NW.

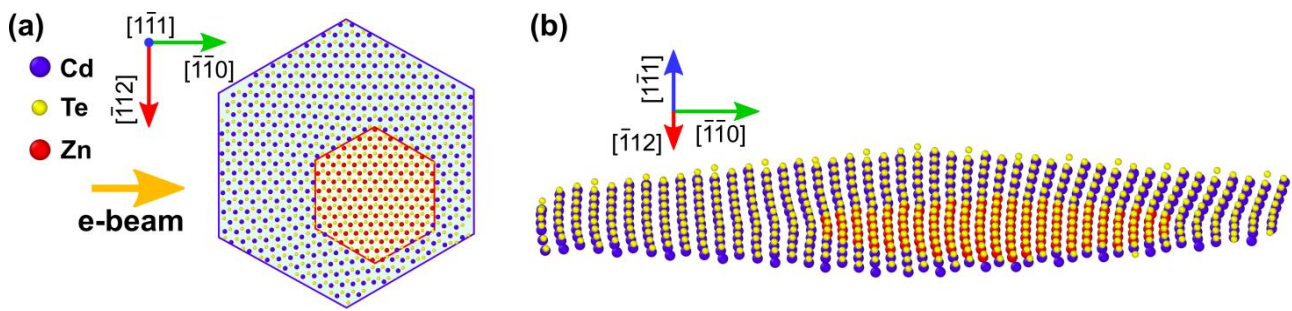

**Figure 7.** Simulation results of the crystal lattice performed in the frame of the finite element method (FEM) for core-shell NW with diameter ~12nm with the a displacements multiplied by the factor of four for clarity reasons. (a) the crystalline model of a cross-section slice across a strained core-shell NW, (b) slice shown in (a) tilted toward [1$\bar{1}$1] direction

For the purposes of the proper interpretation of $\alpha_t$ map and for the understanding of the crystal lattice distortion, the crystalline model of strained core-shell NW was built. The first step was creating solid core-shell NW model and calculation of three-dimensional strain distribution by using COMSOL software. The simulation is performed for a NW with a relatively small diameter of about ~12 nm so that the size of the simulated data is not too large, and allows us to observe the bending of crystal planes into the NW. In the next step, the crystalline model was built using FEM-simulated results in self-made software. The displacement field components ($u_x$, $u_y$, $u_z$) are multiplied by the factor of four for better visibility of the lattice distortion in the crystalline model. A thin NW cross-section slice resulting from the model calculations is shown in Fig. 7. For a better visibility of the crystal plane bending the same slice is presented from two different angles of view: from the top, Fig. 7(a), and after tilting, Fig. 7(b). These calculations confirm that $\alpha_t$ value depends on the radial position on the bent asymmetric core-shell NW. Accordingly, the experimental $\alpha_t$ map (Fig. 6(c)) represents a projection of complex 3D lattice distortion of the NW. Obtained results confirm qualitatively the nonlinear behavior of the angle of twist on the $\alpha_t$ map.

**6. FEM simulation**

The experimental data of strain and orientation mapping are used for 3D reconstruction of the NW geometry and 3D strain distribution. The simulation of the strain distribution is performed based on the linear elastic approach. Moreover, it is performed in the frame of continuum mechanics approximation neglecting the discrete nature of crystalline structure.

For the scopes of our simulations the total radius of the core-shell NW and the elemental composition of the core, which is pure ZnTe, are fixed. At the beginning stage, elemental composition of the shell equal to $Cd_{0.6}Zn_{0.4}Te$ determined from TEM study was also fixed in the model. However, as will be shown below, the elemental shell composition can be found by the fitting procedure. The core taper in the model is chosen to be the same as for the NW determined experimentally from TEM data. The radius of the core, *r*, and its relative position with respect to the center of NW (x,y) are parameters that can be found by fitting the experimental strain and orientation maps to the model calculations. The fitting procedure starts with the considerations of in-plane $\alpha_\parallel$ and out-of-plane $\alpha_\perp$ bending only. We find, in particular, that the core radius, r, and the core position (x,y) with respect to the NW center determine unambiguously the specific NW bending for a given Cd concentration in the shell, Eq.(4).

$$r_i = f(x_i, y_i) \qquad (4)$$

Therefore, for each shell composition (Cd content was chosen in the range 0.4-0.8) a set of core radii with corresponding (x,y) positions, that cause the specific NW bending are obtained. The details of the determination procedure of (x,y) positions of the core are described in the ESM Fig. S2. In the next step, the comparison of experimental to FEM simulated projected strain maps allows us to find a suitable core radius of the NW. In our case, the radial strain map $\varepsilon_r$, Fig. 2(f), which contains region of a sharp strain gradient, is sufficiently characteristic to perform this procedure. It is established that according to the fitting procedure the elemental composition of the shell can also be determined. (This is due to the fact that the maximal value and behavior of strain on radial $\varepsilon_r$ and axial $\varepsilon_a$ maps depends on content of cadmium in the shell, radius and position of the core.) Cd-content in (Cd,Zn)Te shell determined by FEM simulation amounts to 0.57±0.05 which is in good agreement with the EDX data.

Based on FEM modelling, the best fit is obtained for the geometric model in which ZnTe core is fully covered by (Cd,Zn)Te shell. However, this result does not explain well the experimental $\varepsilon_r$ strain map for which only one core-shell transition region is observed, Fig. 2(f).

This effect can be explained by the presence of 4-5 nm thick amorphous layer, which covers the entire NW (Fig. 1(b)). This layer is not visible on the strain maps obtained by NBED and the elastic parameters of this shell are unknown. We associate this amorphous layer to the surface oxidation of the NW. Amorphization of the thin (Cd,Zn)Te shell covering ZnTe core on the right side of the studied NW could be the reason for the absence of the second strain jump (closer to NW surface) on experimental data. For the simplicity reasons, this layer is not included in the FEM model. However, it may explain the fact that the simulated data obtained for regions close to the NW surface may not be comparable with the experiment due to the insensitivity of NBED to the presence of the amorphous layer.

The geometric values obtained by the fitting procedure are: the core radius of 17±1 nm at the bottom part of the NW, the NW core position with respect to the NW center, x=−4.5±0.5 nm, y=8.7±0.5 nm. Importantly, the performance of the above described fitting procedure makes it possible to reconstruct geometry, 3D strain field and orientation for the studied NW.

The results of FEM simulation are presented in Fig. 8. It is found that stresses with the values up to a few gigapascals may occur within the NW Fig. 8(b). The stress along with the asymmetric core-shell geometry causes significant bending of the NW. Maps of in-plane $\alpha_\parallel$ and out-of-plane $\alpha_\perp$ bending components (Figs. 8(c, d)) are in good agreement with experimental ones (Fig. 2(h), 6(b)).

The core-shell geometry induces a complex three-dimensional strain distribution, the projection of which we observe on the experimental strain maps. The calculated projections of radial and axial strain, based on FEM simulated 3D strain distribution, are presented in Fig. 8(e,f). The simulated data confirms the presence sharp strain jump on experimental $\varepsilon_r^{rel}$ map and a smooth strain change for $\varepsilon_a^{rel}$ in the radial direction. The behavior of $\varepsilon_a^{rel}$ map confirms that NW's shell and core crystal structures are coherent in the axial direction. The observed axial strain change is caused by the bending of the NW. Simultaneously, $\varepsilon_r$ jump indicates the transition from the core to the shell region on the map. The presence of a somewhat smoothed jump on the experimental map may be explained by the presence of a topological variation, such as roughness, of the NW core.

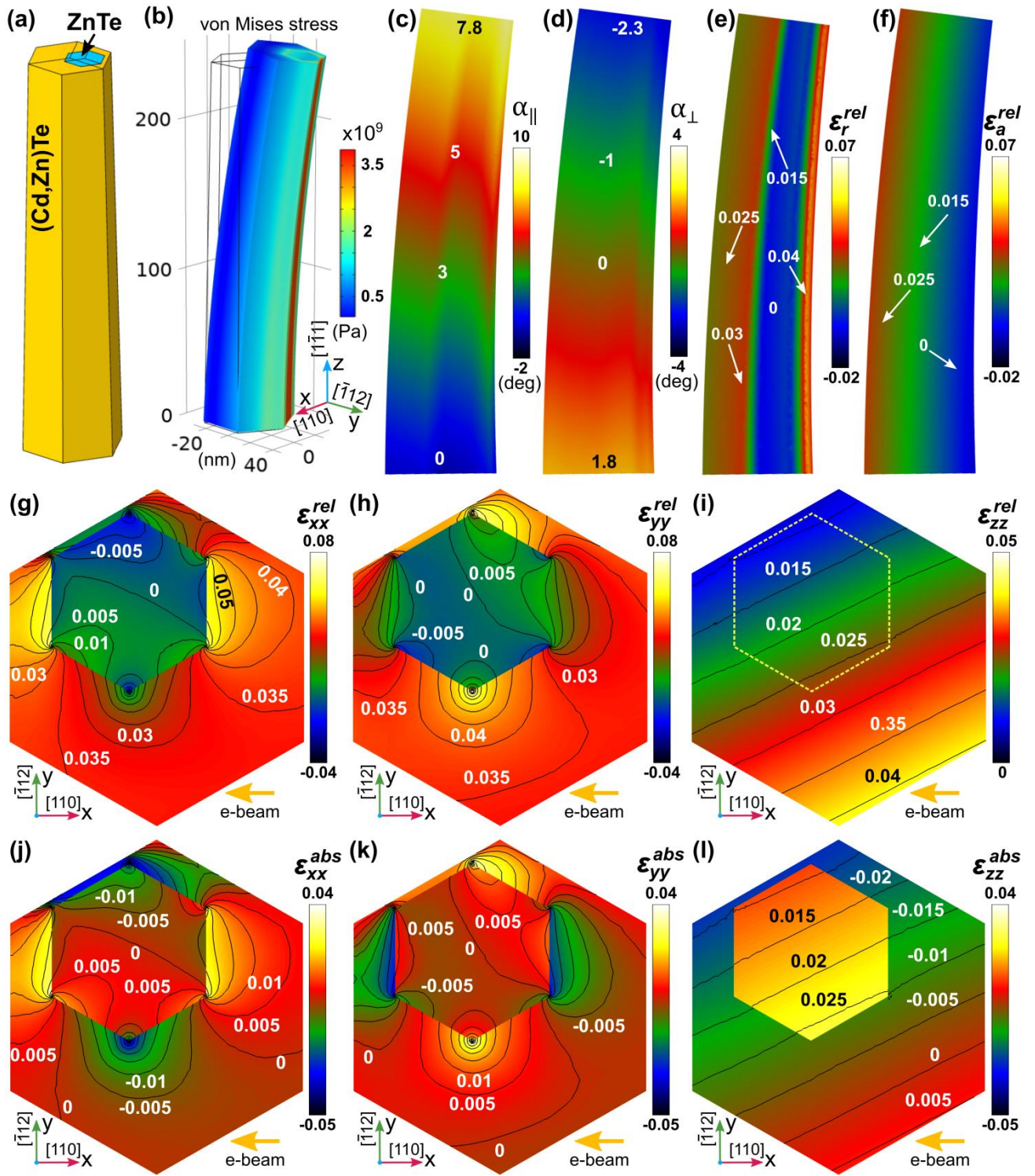

**Figure 8** FEM simulation results. (a) The initial model of asymmetric (Cd,Zn)Te/ZnTe core-shell NW. (b) FEM calculated von Mises stress. (c, d) – maps of in-plane $\alpha_\parallel$ and out-of-plane $\alpha_\perp$ components of the NW's bending. (e, f) – normalized radial $\varepsilon_r^{rel}$ and axial $\varepsilon_a^{rel}$ strain projection maps respectively. (g–i) – calculated relative strain in terms of ZnTe bulk value. (j–l) calculated absolute strain in terms of relaxed ZnTe core and $Cd_{0.57}Zn_{0.43}Te$ shell values.

In addition to projection maps, 3D FEM modeling gives us an opportunity to look inside NW heterostructure and receive information concerning hidden dimensions that cannot be directly observed by single zone axis TEM data. For example, a cross-sectional strain distribution is useful for a comprehensive study of the strain behavior in core-shell NW. The cross-section maps of relative strain calculated using relaxed ZnTe as a reference, at 150 nm NW height, are presented in Fig. 8(g-i). This strain representation allows for tracking the tendency of the lattice changes in the NW. Significant lattice changes were observed on each of the relative strain map. $\varepsilon_{xx}^{rel}$ and $\varepsilon_{yy}^{rel}$ map, which corresponds to radial strain, demonstrates the complex behavior of the lattice change. According to $\varepsilon_{zz}^{rel}$ map the core and shell crystal lattices are coherent in the z (growth) direction, whereas the gradual strain change is caused by the NW bending.

At the same time, the absolute strain (Fig. 8(j-l)) is advantageous for the determination electrical and optical properties of the object. Regarding to 3D FEM results the core undergoes smaller radial deformation compared to the shell (Fig. 8(j,k)). At the same time, ZnTe core is under a significant axial tension (Fig. 8(l)).

## 5. Summary

In conclusion, we developed an approach for strain and orientation mapping of a single nano-object containing twin boundaries and a large thickness gradient. This has been achieved by a comprehensive study of scanning nanobeam electron diffraction data accompanied by a theoretical simulations and the pattern matching technique. The proposed approach is implemented for precise strain and orientation mapping with an ultimate spatial resolution of 2.82×3.91 nm pixel size in our case, however, can be easily improved to electron probe size – of about 1.3 nm, and angular resolution of about 0.1 degrees for curved (Cd,Zn)Te/ZnTe core-shell NW.

Using for acquisition an ordinary NBED data (without beam precession) has enabled us to conserve also dynamical effects on diffraction patterns, which are used for the estimation of local thickness of the object. However, performing a more accurate thickness mapping would require a larger electron probe convergence angles.

It has been experimentally established and confirmed by finite element modeling that asymmetric core-shell geometry introduces complex behaviors of lattice distortion, which appears as a twist component of lattice bending on the projected experimental maps. Obtained strain and crystallographic orientation data are introduced into the finite element modeling for the reconstruction of 3D strain distribution in the NW. The proposed approach can be used for performing low-dose quasi-tomography experiment on strained core-shell nanowires (for example in the case of beam-sensitive materials or if tilting the sample in a holder is limited).

**Acknowledgment**

This work was partly supported by the Polish National Science Centre through grants No. 2016/21/B/ST5/03411, 2017/26/E/ST3/00253, and UMO-2019/35/B/ST5/03434.

**Electronic Supplementary Material:** Supplementary material (FEM modeling of relative tetragonal distortion for (Cd,Zn)Te/ZnTe core-shell NW, and the details of the determination procedure of (x,y) positions of the ZnTe core) is available in the online version of this article at http://dx.doi.org/********.

Notes: The authors declare no competing financial interest.

# Graphical Table of Contents

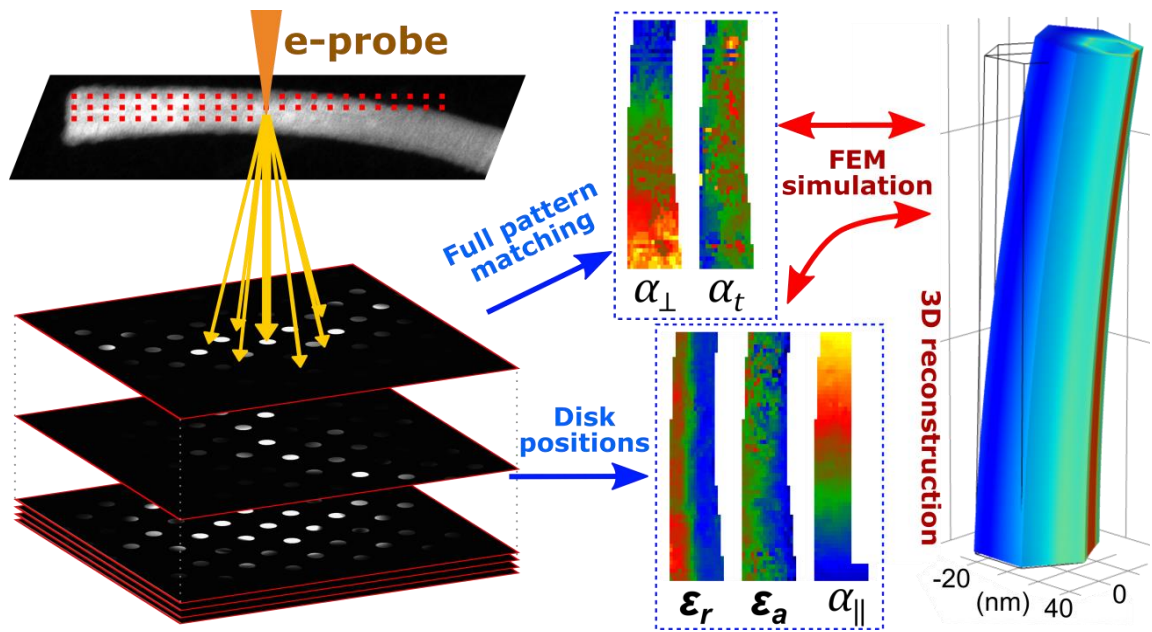

This study concerns the development of a method for strain and orientation mapping of strained highly-asymmetric core-shell nanowire based on transmission electron microscopy data involving single zone axis diffraction pattern. An approach is based on the symbiosis of comprehensive analysis of scanning nanobeam electron diffraction data and finite element method simulation which allow for a full reconstruction of the three-dimensional strain field in the nanowire.

# Electronic Supplementary Material

# Reconstruction of three-dimensional strain field in an asymmetrical curved core-shell hetero-nanowire

Serhii Kryvyi[1,*], Slawomir Kret[1] and Piotr Wojnar[1]

[1]Institute of Physics Polish Academy of Sciences, al. Lotnikow 32/46, 02-668 Warsaw, Poland

E-mail: kryvyi@ifpan.edu.pl

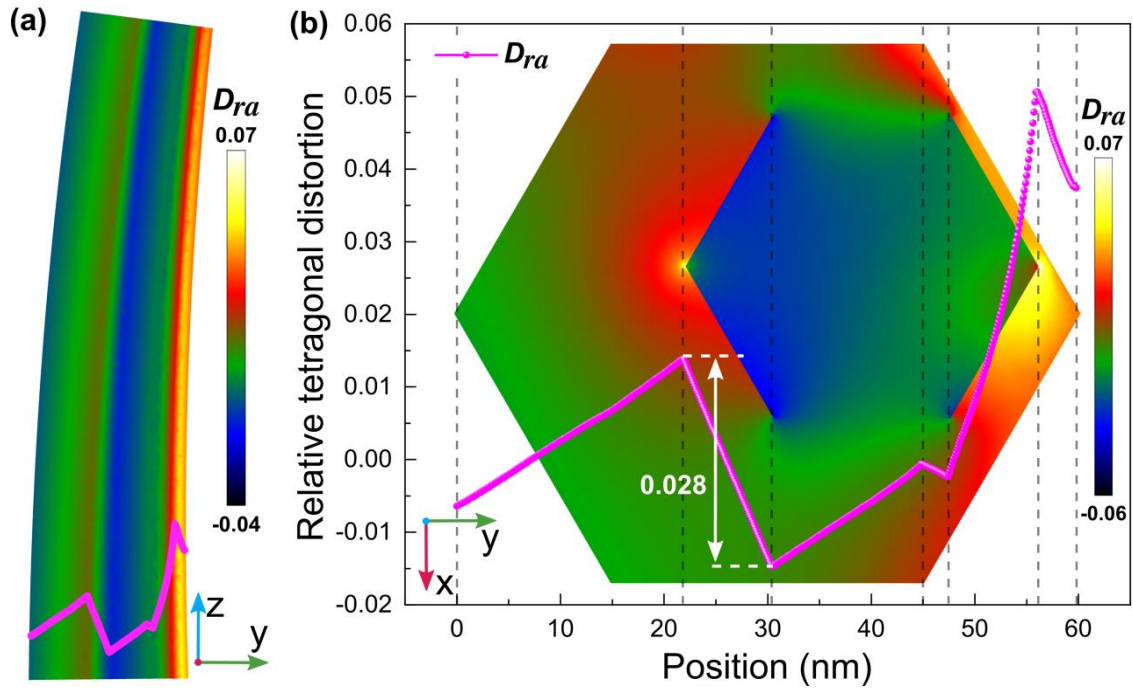

**Fig. S1** Relative tetragonal distortion for (Cd,Zn)Te/ZnTe core-shell NW calculated by FEM. (a) Projection of tetragonal distortion on *yz* plane (perpendicular to electron beam). The pink line represents the distortion profile across the NW. (b) Tetragonal distortion profile across the NW. As the background was set a cross-section (*xy* cut) of tetragonal distortion data. Clearly visible that geometric features of the NW cause extremums on the tetragonal distortion profile.

In Fig. S1, FEM calculated data of relative tetragonal distortion for (Cd,Zn)Te/ZnTe core-shell NW according to Eq. 3 is presented. In order to compare simulated and experimental data, the radial and axial strain components, $\varepsilon_{yy}$ and $\varepsilon_{zz}$, were chosen respectively. Both simulated $\varepsilon_{yy}$ and $\varepsilon_{zz}$ strain components were calculated with respect to ZnTe bulk value. As shown on Fig. S1b, the tetragonal distortion profile reflects geometric features of the NW (in particular the core position). The calculated tetragonal distortion profile (Fig. S1b) agrees well with the experimental one (Fig. 3b). The range of lattice distortion in the middle part of the FEM simulated profile (transition core-shell region) reaches 2.8%. This value is somewhat larger than obtained in the experiment (1.6%), which may be explained by surface roughness and smaller sharpness of ZnTe core corners in the studied NW.

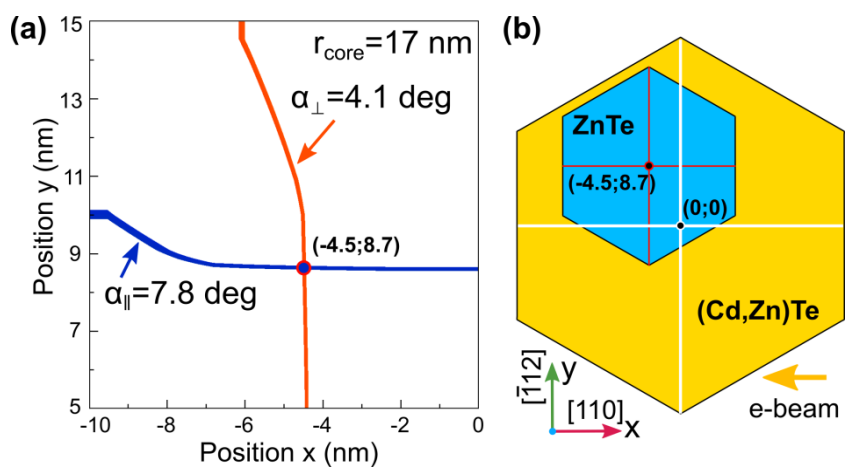

**Fig. S2** The determination of the core position (x,y) within the NW. (a) – isolines showing the relationship between the core position of the radius of 17 nm in the NW and bent angles $\alpha_\parallel$ (blue line) and $\alpha_\perp$ (red line). Those lines' intersection corresponds to the bent NW's core position. (b) – the scheme of the asymmetric core-shell NW.

The procedure of determining the core position is based on finding two isolines on the (x,y) plot, that correspond to the bending of the NW by $\alpha_\parallel$=7.8 degrees and $\alpha_\perp$=4.1 degrees respectively (Fig. S2a). The intersection of those lines provides information on the exact core position (x,y). This position depends on both Cd-content in $Cd_xZn_{1-x}Te$ shell and core radius. As a result, the core positions were determined for each simulated Cd-content and the core radius. In the next step, the comparison of experimental to FEM simulated projected strain maps allows us to find a suitable core radius and Cd-content of the NW.